# Stock Price Prediction using Dynamic Neural Networks

David Noel

Department of Electrical and Computer Engineering, University of Florida

**Abstract**

*This paper will analyze and implement a time series dynamic neural network to predict daily closing stock prices. Neural networks possess unsurpassed abilities in identifying underlying patterns in chaotic, non-linear, and seemingly random data, thus providing a mechanism to predict stock price movements much more precisely than many current techniques. Contemporary methods for stock analysis, including fundamental, technical, and regression techniques, are conversed and paralleled with the performance of neural networks. Also, the Efficient Market Hypothesis (EMH) is presented and contrasted with Chaos theory using neural networks. This paper will refute the EMH and support Chaos theory. Finally, recommendations for using neural networks in stock price prediction will be presented.*

## 1. INTRODUCTION

THIS paper uses a Dynamic Neural Network to predict a given stock's future daily closing price. These are models used for predictions and non-linear filtering by analyzing data for patterns and "learning" from said patterns. Two prevailing theories seek to qualify market behavior; EMH and Chaos.

The **EMH theory** states that no system can reliably 'beat' the market because a stock's price has already incorporated and reflects all factors into its valuation, thus being *random* and *unpredictable*. In contrast, the contrary **Chaos theory** states that stock prices are made up of *deterministic* plus *random* patterns. [1]

Chaos represents deterministic but non-linear processes that appear random because they cannot be easily quantified. Through the neural networks' inherent ability to learn non-linear, dynamic, & chaotic systems, it is possible to outperform traditional stock price analysis; much to the liking of investors who can exploit this ability for huge profits.

Neural networks enable multivariate models, allowing parallel processing of input data, thus quickly processing large amounts of data. Their primary strength is that they can determine patterns and anomalies within data, but most importantly, their ability to detect multidimensional & non-linear connections within that data; this makes neural networks extremely suitable for the dynamic and non-linear data generated by the stock market [2].

## 2. RELATED WORKS

Several fundamental and technical indicators have been adopted for stock price predictions but with inconsistent results [2]. No single method or combination has succeeded enough to outperform the market reliably.

These time series predictions were linear parametric autoregressive, auto-regressive moving averages or moving averages in nature based on principles by Box and Jenkins. Due to their linear nature, these models failed to capture non-stationary/dynamic signals properly.

The use of technical indicators stems from the idea that stock prices move in trends governed by investors' emotions, views, and attitudes due to company news or macroeconomic data. Using data such as volume, momentum, opening/closing prices, and oversold/overbought conditions, technicians use many charts to predict a stock's future value or index under the *belief* that history repeats itself. This methodology contradicts the EMH hypothesis but is used by over 80% of stock traders [5]. The major downside of technical analysis is that it is *highly subjective* since each trader can interpret the charts differently, making this method inconsistent, biased, and maybe most appropriate for *very short-term* predictions.

Fundamental analysis involves a more in-depth study of the company. Fundamental analysts study the company's financial performance, price/earnings, consumer demand, overall market conditions, and competitors to determine the *intrinsic value* of a stock. This value indicates what a particular stock is "truly" worth, and they believe that the market will eventually realize this intrinsic value and reward the stock by maintaining or exceeding the said price. This method, however, involves processing huge amounts of data and is subject to individual interpretation. Thus, even if the data suggests a favorable movement in the stock, it may take a long time for the rest of the market to realize or catch up. This method, thus, is suitable for *long-term predictions* and growth.

The algorithms, neural network architectures, and associated functions will be implemented using MATLAB® and its powerful Neural Network Toolbox.

## 3. ALGORITHM

The **Levenberg-Marquardt (LM)** algorithm will be used for training purposes. This algorithm quickly updates weight and bias values using back-propagation techniques, and it seems to be the quickest method for training reasonable-sized feedforward neural networks (up to weights reaching several hundred). [3]

The design of the algorithm was such that it approaches 2<sup>nd</sup> order training speed without computing the Hessian matrix (unlike the Gauss-Newton method).

In the LM algorithm, a function $f(x)$ is minimized to a sum of squares, i.e.

$$\min_{x} f(x) = \|F(x)\|_2^2 = \sum_i F_i^2(x) \tag{1}$$

Assuming output, $y(x, t)$, for some continuous model trajectory, $\varphi(t)$, at vector $x$ and scalar $t$, we have:

$$\min_{x \in \mathfrak{R}^n} \int_{t_1}^{t_2} (y(x,t) - \varphi(t))^2 dt, \quad (2)$$

where $y(x,t)$ and $\varphi(t)$ are scalar functions. Discretizing (2) using the appropriate quadrature formula, we get a least squares problem:

$$\min_{x \in \mathfrak{R}^n} f(x) = \sum_{i=1}^{m} (\bar{y}(x,t_i) - \bar{\varphi}(t_i))^2, \quad (3)$$

where $\bar{y}$ and $\bar{\varphi}$ include the weights. Now from (1),

$$F(x) = \begin{bmatrix} \bar{y}(x,t_1) - \bar{\varphi}(t_1) \\ \bar{y}(x,t_2) - \bar{\varphi}(t_2) \\ \dots \\ \bar{y}(x,t_m) - \bar{\varphi}(t_m) \end{bmatrix}. \quad (4)$$

*Denoting:*
Jacobian $F(x)$ as $J(x)$,
gradient vector of $f(x)$ as $G(x)$,
Hessian matrix of $f(x)$ as $H(x)$,
Hessian matrix of $F_i(x)$ as $H_i(x)$,

*we get,*

$$\begin{aligned} G(x) &= 2J(x)^T F(x) \\ H(x) &= 2J(x)^T J(x) + 2Q(x), \end{aligned} \quad (5)$$

*where,*

$$Q(x) = \sum_{i=1}^{m} F_i(x) \cdot H_i(x). \quad (6)$$

as $x_k$ approaches the solution, $F(x) \to 0$ & $Q(x) \to 0$.

Now as $Q(x) \to \infty$, the Gauss-Newton method experiences problems resolved by the LM method. In such cases, the LM method uses a search direction, $d_k$, that is a solution to the set of linear equations:

$$\left(J(x_k)^T J(x_k) + \lambda_k I\right) d_k = -J(x_k)^T F(x_k), \quad (7)$$

where, $\lambda_k$ dictates the magnitude and direction of $d_k$

The LM method thus uses a search direction that is a hybrid between the Gauss-Newton and the steepest descent direction. The neural models used for training in this paper would adopt a series-parallel (open-loop) network since training adopts "true" output values. For this model, the prediction equations used to determine the output $y(k + 1)$ uses past inputs of $u_1(k), u_1(k-1), \dots, u_1(k-d_u)$ for the 1st time sequence and $u_2(k), u_2(k-1), \dots, u_2(k-d_u)$ for the second time sequence and past outputs(targets) of $y(k), y(k-1), \dots, y(k-d_y)$, all as input data. This can be written in the form:

$$y(k+1) = \Phi_o \left\{ w_{b0} + \sum_{h=1}^{N} w_{ho} \cdot \Phi_h(w_{ho} + \sum_{i1=0}^{d_{u1}} w_{i1h} u_1(k-i1) + \sum_{i2=0}^{d_{u2}} w_{i2h} u_2(k-i2) + \sum_{j=0}^{d_y} w_{jh} \cdot y(k-j)) \right\} \quad (8)$$

and depicted as shown in Figure 1 below:

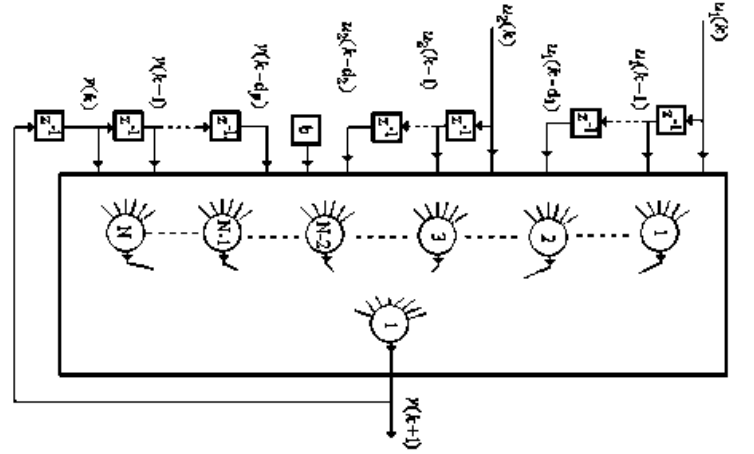

**Figure 1.** NARX(series-parallel) with 2 tapped delay lines for two-time series inputs.

The NARX network, however, is not without issues, and one of them has to do with overtraining caused by the number of connections and weights within the network – the number of parameters. Overtraining can lead to over-fitting, defeating the neural network's purpose. To overcome this issue, the NARX network integrates a regularization technique into its performance function to reduce the number of parameters, thereby reducing its weight and biases. This new function, known as MSEREG, is described as follows:

$$MSE = \frac{1}{N} \sum_{i=1}^{N} (e_i)^2 = \frac{1}{N} \sum_{i=1}^{N} (t_i - y_i)^2 \quad (9)$$

$$MSW = \frac{1}{n} \sum_{j=1}^{n} w_j^2 \quad (10)$$

$$MSEreg = \xi MSE + (1 - \xi) MSW \quad (11)$$

where $\xi$ is the performance ratio and $t_i$ is the target.

## 4. NETWORK

### 4.1 Architecture

The architectural approach taken in this paper for stock price predictions is the "Non-linear Autoregressive with exogenous input (NARX model)" [6]. This powerful network has demonstrated great accuracy and flexibility in modeling non-linear time series data, such as stock data. Using NARX, the idea is to predict future values of a time series $y(t)$ from past values of said time series coupled with past values of a second time series $x(t)$, which can be written as follows,

$$y(t) = f(y(t-1), \ldots, y(t-d), x(t-1), \ldots, (t-d))$$

The NARX network is a two-layer feedforward network consisting of a sigmoid transfer function in the hidden layer and a linear transfer function in the output layer. The network also makes use of tapped delay lines to store previous values of the $x(t)$ and $y(t)$ sequences. It is important to note here that the output, $y(t)$, is then fed back as input to the network, which is needed in the dynamic nature of stock price prediction.

The architecture's open-loop (series-parallel) implementation, as shown in Figure 2(a), will be used to train the network since the 'true' output, instead of the estimated output, is used as input feedback. This has several advantages, the first of which is that the input to the feedforward network is more accurate, and second, the resulting network has a pure feedforward architecture, and thus a more efficient algorithm( static back-propagation )can be used for training.

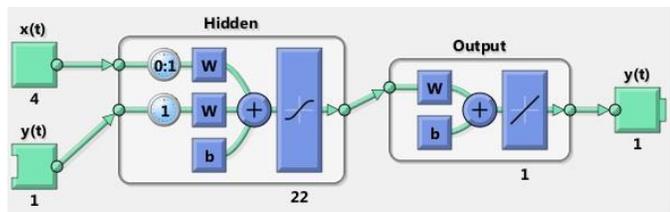

**Figure 2(a).** Open-loop used for training.

Once the network is reasonably trained, it will be converted to a closed-loop network to perform multi-step-ahead predictions. While open, the network was predicting the next value of *y(t)* from the previous values of *y(t)* and *x(t)*. When the network is closed, it can perform multi-step-ahead predictions. This happens because *predictions of $y(t)$* will be used in place of *actual future values of $y(t)$*. i.e., a closed-loop network outputs $y(t+1)$ instead of $y(t)$, as shown in Figure 2(b) below.

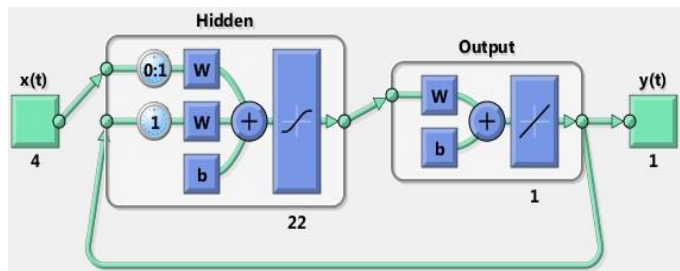

**Figure 1(b).** Closed-loop network for multi-step-ahead predictions.

### 4.2 Data Acquisition

Four years (01/01/2010 – 03/31/2014) of prior stock data would be retrieved from *finance.yahoo.com* using a custom-written function based on MATLAB data-feed plugin. This function automatically segments the data into *input, target, and sample* sequences. The data would be in the format *[date, opening price, high price, low price, volume, closing price, adjusted close price]* and prepared for input to the network using MATLAB preparets function. This function automatically shifts time series data (input and target) by as many steps as needed to fill the initial delay states, thereby greatly reducing errors and complexity. Table I shows the sample dataset used.

**Table I**
*Sample Dataset*

| Timestep(day) | 734506 | 734507 | 734508 | 734509 | 734510 |
|---|---|---|---|---|---|
| Open(Input) | 21.01 | 21.12 | 21.19 | 20.67 | 20.71 |
| High | 21.05 | 21.2 | 21.21 | 20.82 | 20.77 |
| Low | 20.78 | 21.05 | 20.9 | 20.55 | 20.27 |
| Volume | 58223800 | 75206200 | 61810500 | 116669000 | 74806100 |
| Close(Target) | 20.85 | 21.15 | 20.94 | 20.77 | 20.66 |
| Adjusted Close | 18.54 | 18.8 | 18.62 | 18.47 | 18.37 |

### 4.3 Training, Testing, and Validation

As shown in Figure 1, the open-loop network will train the network using tapped delay lines with 2 delays for both the input and output. This will allow training to begin with the 3$^{rd}$ data point. The open-loop network takes 2 input sequences, $x(t), y(t)$, where $x(t)$ is the input signal and is composed of the: *{open price, high price, low price, volume}* data for the stock being analyzed.

It is important to note that for this network, $y(t)$, which is the *closing price* of the stock, is a feedback signal and is also an input and an output (target). Furthermore, 70% of the input dataset is used for training, 15% for validation, and 15% for testing. The network was retrained 10 times to ensure maximum accuracy, using the MSEREG performance function to improve generalizations. See Figures 3 and 4 below for further performance and training parameters used.

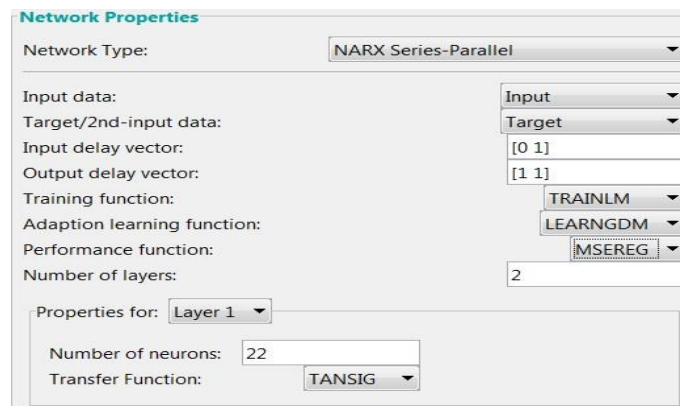

**Figure 3.** Network properties used.

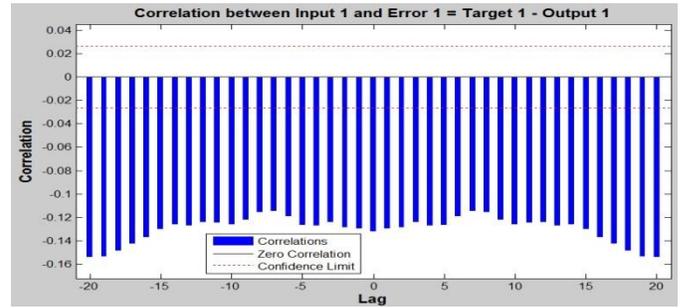

**Figure 4.** Training parameters used.

To determine the best possible parameters for achieving a robustly trained network, the number of input and output delays and the number of neurons were changed systematically, and each combination's RSE and performance ratio were recorded. Table II shows the recorded data after several trials, and the best number of delays and neurons were selected based on depicted performance criteria.

**Table II**
*Performance Based on Input Delays, Target Delays, and Neurons*

| Input delays($d_u$) | Feedback delays($d_y$) | Neurons (N) | Performance | RSE |
|---|---|---|---|---|
| 2:5 | 1 | 3 | 0.016 | 0.99852 |
| 0:1 | 1 | 5 | 0.0154 | 0.99846 |
| 0:1 | 1 | 10 | 0.0156 | 0.99844 |
| 2:5 | 1 | 10 | 0.024 | 0.99844 |
| 6:25 | 1 | 10 | 0.0354 | 0.99849 |
| 0:5 | 1 | 14 | 0.0209 | 0.99847 |
| 0:5 | 1 | 16 | 0.0168 | 0.99862 |
| 0:1 | 1 | 21 | 0.0217 | 0.9954 |
| 26:30 | 1 | 10 | 0.0955 | 0.9977 |
| 0:5 | 1 | 16 | 0.0254 | 0.9983 |
| 0:5 | 1 | 18 | 0.038 | 0.99846 |
| 0:1 | 1 | 22 | 0.0178 | 0.99857 |
| 0:1 | 1 | 23 | 0.0182 | 0.99866 |
| 0:1 | 1 | 25 | 0.01599 | 0.99876 |
| 0:1 | 1 | 30 | 0.0466 | 0.99766 |

*Pink – Out of Confidence Level, *Green – Within Confidence Level
*(0:1) – means integer values 0 to 1 inclusive*

One neuron was used in the feedback layer for all trials because our target is only 1 value. From the data in Table II, it could be seen that when the input delays were between 0 and 25 and the number of neurons was less than 21, the performance and RSE values indicated a very good fit between the actual and predicted data (the *pink portion of the chart*). However, it was observed that in those cases, the Input-Error correlation curves were significantly out of the confidence levels, as can be seen in Figure 5.

**Figure 5.** Input-Error correlation out of confidence limit with input delays 2:25 and ≤ 21 neurons.

After further testing (*green portion of Table II*), it was observed that the performance and accuracy increased with delays above 25 and 10 neurons or less. However, this large number of delays gave a performance hit to the network and was not chosen as appropriate input. Throughout the trials, it was observed that input delays between 0 and 1 yielded the BEST results across all performance metrics when the number of neurons was between 22 and 30 inclusive. The best performance was seen using a minimum of 22 neurons and [0:1] input delays, as depicted in Figure 6 below.

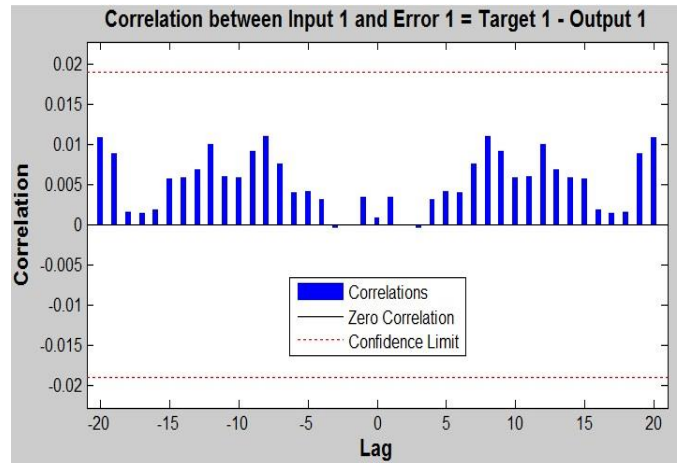

**Figure 6.** Best performance seen at $d_u$=0:1, $d_y$=1, N=22

As such, the network used for final training will use the parameter values depicted in Figures 3 and 4 above, with the Levenberg-Marquardt (LM) algorithm, 2 hidden layers, and 22 neurons.

### 4.4 Model Acceptance Criteria

A Mean Squared Error (MSE) of Zero means no errors, and a Regression (R) value of 1 means a 'perfect' relationship between the input and output data. It measures how well the targets describe the variation in the output. The acceptance criteria would be MSE values close to Zero and R-Values close to 1. Results are further verified using correlation, performance, and time-series charts. Divergences greater than 10% of the nominal values would prompt retraining in which the number of neurons or delays would be changed. [7]

# 5. RESULTS

The model was first trained on 4 years of stock data (01/01/2010 – 1/01/2014) using the series-parallel model shown in Figure 1, after which 100 days of simulation was conducted to predict the closing prices, using the parallel(closed) network as seen in Figure 2(b).

## 5.1 Training

**Table III**
*Performance Results – Trained Network*

| Stock | Training Perf. | MSE | RSE(ALL) | Divergence(max) |
|---|---|---|---|---|
| Intel (INTC) | 0.0147 | 0.024288 | 0.998 | 1.122% |

Table III shows the performance of the *open* network. It meets our acceptance criteria with an RSE of 0.998 and an MSE of 0.0242. Its performance of 0.0147 indicates it has very high efficiency and is much better than any linear-modeled dynamic system for predicting stock prices [5].

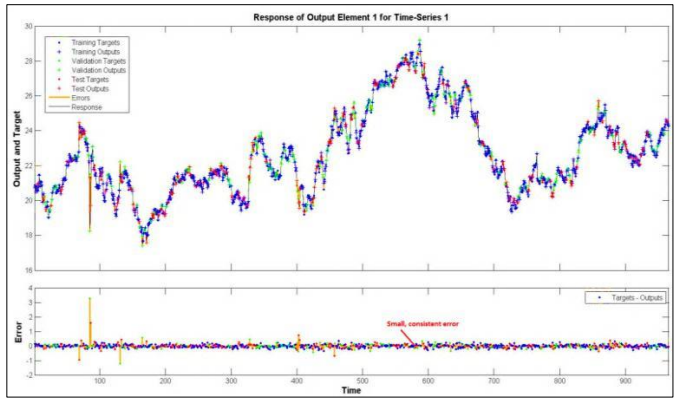

**Figure 7.** Correlations between training, validation, test, and error.

Figure 7 depicts the relative performances obtained for training the network. The training targets and outputs, validation and outputs, and testing targets and outputs are extremely close, with less than 2% divergence from the targets. This close correlation is further verified by the Error vs. Time graph shown. This is important because we want the error in the trained network (one-step-ahead) to be very small so that the predictions (multi-step-ahead) made in the parallel network are as accurate as possible.

Figure 8 shows the errors between the "actual" outputs vs. the "predicted" outputs while training.

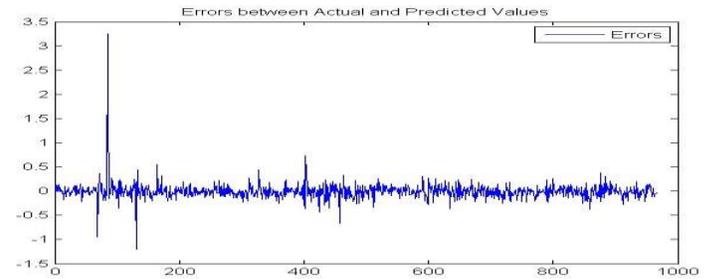

**Figure 8.** Errors between actual and predicted values during training.

This graph clearly shows that the divergences/errors between actual and predicted outputs are small and contained around zero. The maximum divergence from the actual values was approx. 1.122%, which is acceptable for our model. The 5 very large spikes are most likely caused by extremely random events, such as the company unexpectedly reporting a dividend increase or share buy-back, in which the stock rallied.

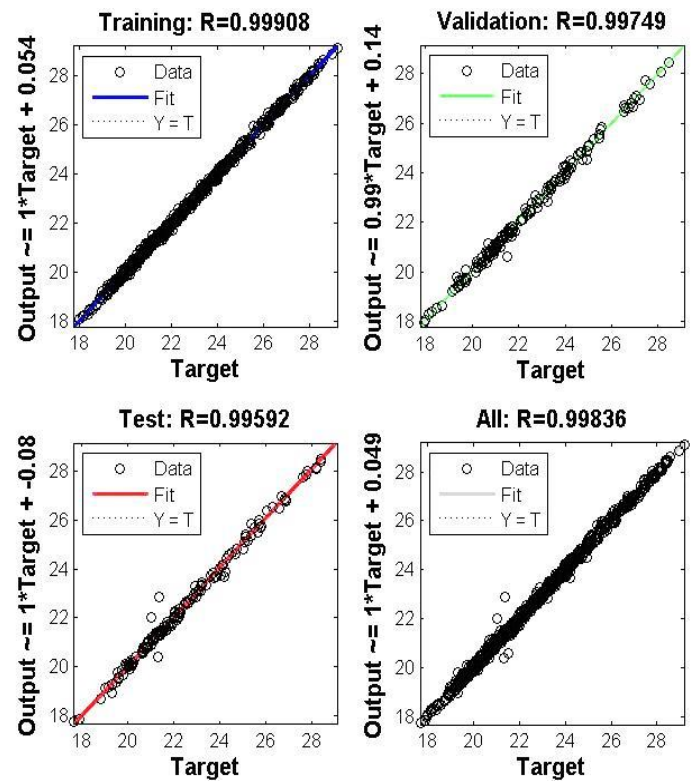

**Figure 9.** Regression curves between output and target values.

The R-values between the outputs and targets measure how well the targets explain the variation in the output. A number equal to 1 signifies a perfect correlation between targets and outputs. In our model, the value is approx. 0.998, which signifies a very accurate correlation.

Figure 9 above shows the input-error cross-correlation function, which depicts how the errors are correlated with the input sequence $x(t)$. In a perfect model, all the correlations should be zero, and if they are not, it signifies room for improvement by adjusting the number of neurons in the hidden layer or the number of delays until the bars fall within the confidence bounds around zero. In our model, the bars all fall within the confidence limits of [-0.018, 0.018]. This further signifies that our model has a very efficient design, thus, will produce great accuracy.

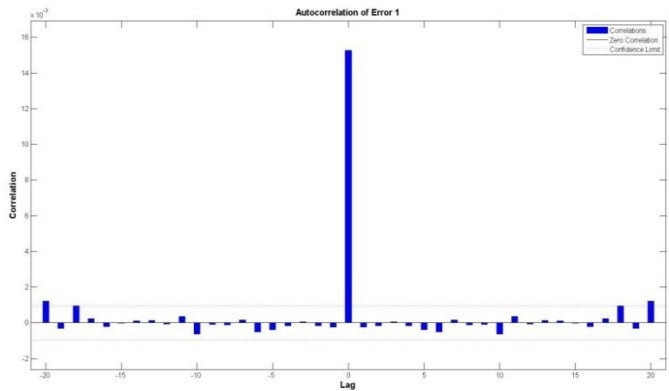

**Figure 10.** Error autocorrelation to validated network performance.

Figure 10 shows the error autocorrelation function, which validates the network performance by showing how the prediction errors are related to time. For a perfect model, there should only be one nonzero value occurring at Lag 0 (the *MSE* of the network), which would signify that prediction errors are only white noise. For our model, all the bars except the one at Lag 0 fall within the 95% confidence limit of around 0, which validates this model as very accurate. The 5% loss in performance accuracy is mainly white noise, which is inherent in the dynamic nature of the stock market.

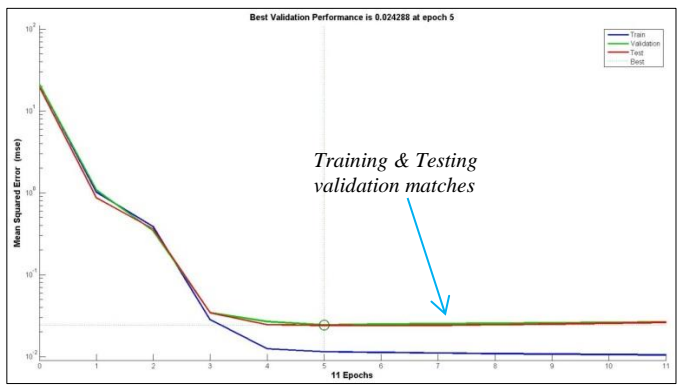

**Figure 11.** Performance plot of training, validation, and testing.

Figure 11 shows plots of performance, validation, and testing during training. The network errors decreased until iteration 5 (out of 11 iterations), where their gradients were no longer changing. Before iteration 5, the gradients of the curves were not increasing. Therefore, it can be concluded that over-fitting did not occur, and this network will produce very accurate results.

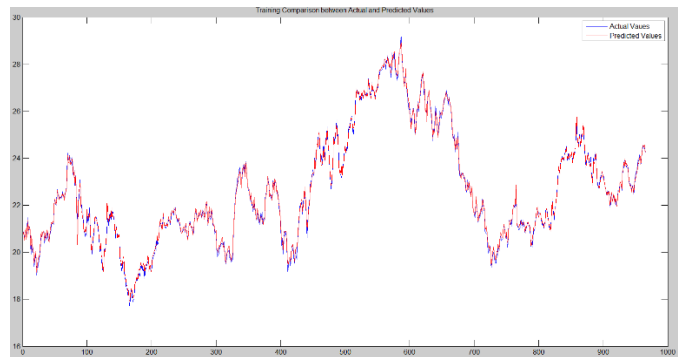

**Figure 12.** Output vs. target during training.

Figure 12 shows the output values obtained during training vs. the target values. From this chart and the previous performance graphs, the network *VERY* accurately models the *closing price* of the stock index, given the four inputs of *opening, high, low* price, and trading *volume*, for the training period of 4 years. ***Red signifies the predicted values,*** while ***blue represents the target values***. The slight deviations are mainly due to the white noise inherent in such data and could be minimized using an appropriate filter.

### 5.2 Simulation and Prediction

Now that training is completed using the series-parallel(open) network with acceptable accuracy, the network will now be converted to a purely parallel(closed) one, as shown in Figure 2(b), and the target prices will be predicted for 100 days, using data from *1/1/2014 – 3/31/2014*.

**Table IV**
*Performance Results for Simulated Network*

| STOCK | Simulated Perf. | MSE | RSE(ALL) | Divergence(max) |
|---|---|---|---|---|
| Intel (INTC) | 0.0142 | 0.024288 | 0.998 | 1.173% |

Table IV shows the performance of the *closed* network. It also meets our acceptance criteria with an RSE of 0.998 and MSE of 0.0242. Its performance of 0.0142 also closely matches the expected open network performance.

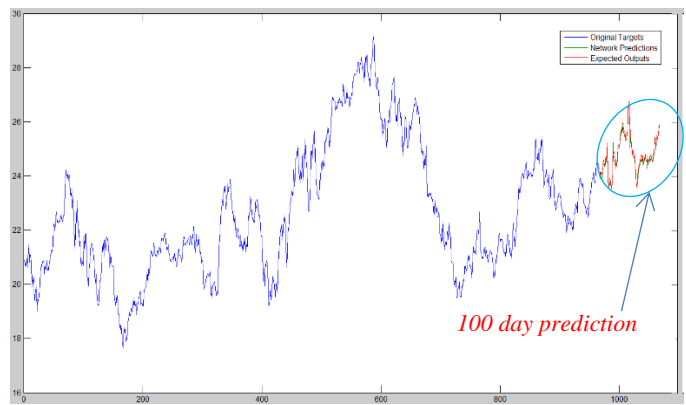

**Figure 13.** Simulated output showing targets and predictions.

Figure 13 shows 3 charts combined. The *blue represents the Original (actual) targets*, the *red shows the expected outputs* after simulation, and the *green shows the network's predictions*. The chart shows that the network predictions and the expected outputs are very close. The network predictions parallel the expected outputs with only a 1.17% maximum divergence.

This accuracy was expected because the trained network showed superior performance characteristics, the same characteristics used in the simulated network. I.e., the simulated network also has an MSE of 0.0242 and an R-Value of 0.998.

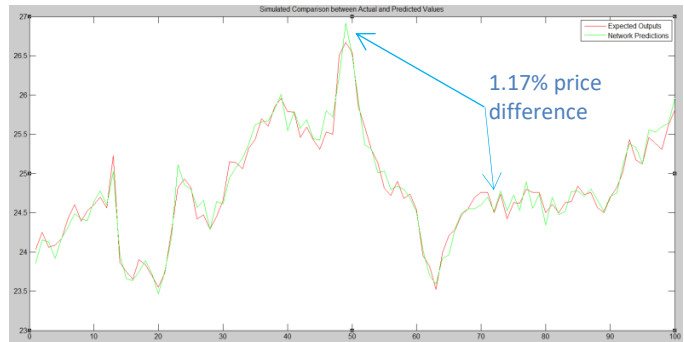

**Figure 14.** Results of the simulated network showing target values vs. predicted values.

Figure 14 is an enlarged version of the circular region identified in Figure 13. It shows the *network predictions in green* and the *target (actual) values in red*. Given the dynamic and random nature of stock prices, this prediction is a very good fit for the actual values, further validating the efficiency of the network model used and the potential of neural networks in general.

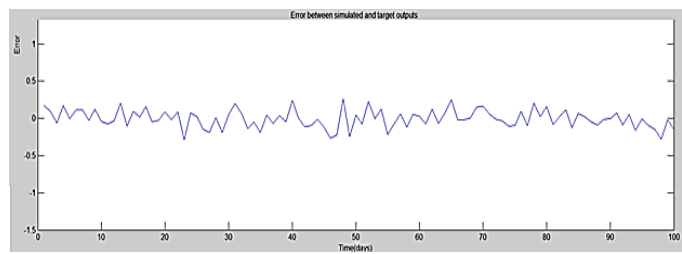

**Figure 15.** Errors between simulated outputs and targets.

Figure 15 shows the errors between the *simulated outputs* and the *target values* for the 100 simulation days using the closed network. It represents the differences between the charts shown in Figure 14. The errors are very small and fluctuate around the zero line. The maximum error difference recorded was 1.17%, indicating that the predicted (simulated) and target (actual) values are sufficiently close to each other to be deemed negligible. Again, most of the error is due to white noise, which can be reduced by first filtering the data. This filtering could be implemented using an extended Kalman filter which would seek to predict random changes in the stock price movement due to company and other macro-economic activities.

## 6. CONCLUSION

From the above data and analysis, it is evident that well-trained neural networks can quickly learn patterns within very large, seemingly unrelated datasets and make intelligent and mostly accurate predictions. It is also clear that neural networks eliminate or minimize randomness even when human counterparts fail. This is most likely because neural networks can quickly recognize and learn extremely subtle patterns in large datasets, which may go unrecognized by humans. This ability to 'learn or model' randomness directly supports the Chaos theory and refutes the EMH theory, and thus can be used to 'beat' the market

## 7. RECOMMENDATIONS

The errors that appeared as white noise could be reduced in several ways. First, the input data could be pre-processed using the techniques of Independent Component Analysis or even another training algorithm, then fed as input to the neural network. Secondly, the input data could be filtered using an extended Kalman filter which also makes intelligent predictions, then fed as input to the network.